# Prevent eavesdropping with bright reference pulses for practical quantum key distribution

Guang Wu, Jie Chen, Yao Li, Lilin Xu, and Heping Zeng[1]

Key Laboratory of Optical and Magnetic Resonance Spectroscopy, and Department of Physics,
East China Normal University, Shanghai 200062, People's Republic of China

Abstract

We analyze the application of bright reference pulses to prevent the photon-number-splitting attack in weak-pulse quantum key distribution. Under the optimal eavesdropping strategy as far as we know, the optimal parameters of bright reference and signal pulses can ensure a secure transmission distance up to 146 km. To realize the quantum key distribution scenario with up-present techniques, we present an experimentally feasible scheme to create a large splitting ratio between bright reference and signal pulses, and to switch the bright reference pulses away from signal pulses to avoid the after-pulse disturbance.

[1] Corresponding author. Electronic address: hpzeng@phy.ecnu.edu.cn.

## I. Introduction

Quantum cryptography (QC) can achieve absolutely secure communications [1], which is guaranteed by the fundamental laws of quantum mechanics [2]. However, the practical QC is limited in distance by the bottleneck of the up-present techniques such as lossy quantum channel, imperfect detectors and single-photon sources. As a result, the game between sender and eavesdropper still exists [3-5]. Thanks to the improvements of single-photon detectors and quantum channel, the experimental quantum key distribution (QKD) has made great progress ever since the first experimental demonstration at only 30-cm distance in 1989 [6]. And now the QKD could be implemented in more than 100 km in single-mode fibers [7, 8]. Almost all these QKD experiments used the weak pulse sources since ideal single-photon sources [9] are still immature for practical QKD applications. Although the security of practical QKDs with weak-pulse sources has been proved [10], the secure distance of these QKDs are highly limited by the photon-number-splitting (PNS) attack.

In the PNS-attack scenario, Eve splits the multi-photon pulses which are unavoidable in weak-pulse source due to the Poissonian statistic feature of the classical source. Then she can take at least one photon and store it in the quantum memory while sending the residual to Bob through her ideal channel. Eve maintains the quantum states until Alice broadcasts her basis information. Then she probes the photons with the same basis according to Alice's broadcasted information, and obtains the information of the multi-photon pulses as the same as Alice. Eve's ideal channel can ensure these multi-photon pulses reach Bob's set-up without attenuation and generate the same key rates as Bob's expectation value without the contribution of single-photon pulses. Eve forces all Bob's information to be generated from the multi-photon pulses by suppressing single-photon pulses, she can thus eavesdrop all the information Bob obtained (from the multi-photon pulses only) without being discovered as all Bob's information is decoded after a lossy channel.

Hwang, Wang, and H. K. Lo put forward decoy states in Ref. [11] that extend the unconditional transmission distance up to 140 km. The decoy states can actively monitor the PNS attacks by analyzing the yields of signal and decoy-state photons. The original scenario requires infinite decoy states, which is difficult for practical applications. Later on, practical decoy-state scenarios were developed [12] on the basis of long-term stable operation to collect enough decoy and signal states. So far, almost all the long-term stable QKDs in long-distance fiber are based on phase encoding, which are difficult to produce multi-decoy states at a high speed. Fortunately, an active feedback technique to auto-compensate unpredictable polarization scrambling enables the control on single-photon pulse polarization for more than 10 hours in long-distance fiber up to 100 km [13]. Experimental results of long-term operations in 50, 75 and 100 km fibers indicate that such a single-photon pulse polarization control supports stable polarization-encoding in long-distance fibers to facilitate stable "one-way" fiber system for polarization-encoded QKD, which can be readily extended for high-speed decoy-state QKD applications.

In this paper, we present a detailed theoretical analysis on a practical QKD scenario with bright reference pulses (BRPs) to prevent against the PNS attack. This paper is organized as follows. After this brief introduction, we introduce the principle of BRPs-QKD scenario and its derivatives in Sec. II. In Sec. III we present an optimal eavesdropping strategy and the optimal parameters for the practical implementation of BRPs-QKD. BRPs-QKD requires a large splitting ratio between bright reference and signal pulses. We discuss a practical solution to meet this requirement in Sec. IV, where we also present an experimentally feasible scheme to switch the bright reference pulses away from signal pulses to avoid the after-pulse disturbance. A brief summary is given in Sec. V.

## II. Bright reference pulses in quantum key distribution

It is simple to beat PNS attack with bright reference pulses (BRPs) in phase-encoded QKD system with weak-pulse sources. The scenario came from the original B92 protocol [14], which aimed to prevent the intercept-resend attacks. Figure 1 presents the typical schematic set-up in a fiber QKD system. Alice uses an unsymmetrical Mach-Zehnder interferometer (MZ1) to produce a BRP followed by a weak signal pulse. Bob uses another unsymmetrical Mach-Zehnder (MZ2) to decode the signals. If the arm length difference of MZ1 is the same as that of MZ2, there are three photon pulses that enter Bob's single-photon detectors successively. The first photon pulse is attenuated by both BS1 and BS2, signed as the dim pulse (D). The second photon pulse is attenuated by BS1 or BS2 which include Alice and Bob's phase information, signed as the signal pulse (S). The third photon pulse encounters no attenuation, signed as the BRP (B). In the ideal condition, Bob can detect the BRP every clock cycle. If Eve wants to suppress both the signal pulses and the BRPs, Bob will detect Eve immediately. Otherwise if Eve only suppresses the signal pulses, the BRPs will produce half interference errors in Bob's MZ interferometer. Alice and Bob can find her by comparing their bit error rate.

The application of BRPs against PNS attacks was first mentioned in Ref. [15]. Shortly, M. Koashi put forward a new scenario with an individual channel to transmit BRPs against PNS attacks [16]. Nevertheless, its practical applications in long-distance fiber are difficult due to the unpredictable phase drifts. Ref. [17] presents a sideband detection scheme that uses BRPs to beat

PNS attacks. The time-bin of signal pulses at the sideband frequencies $\omega_0+\Omega$ and $\omega_0-\Omega$ are the same as that of the BRPs at $\omega_0$. Then the only way to avoid the disturbance of BRPs is to improve the isolation of the filter, which is limited by the up-present techniques. From the experimental points of view, the scenarios of Refs. [16, 17] are difficult for long-distance QKD applications, while the original scenario of BRPs in Ref. [14] is more practicable. In what follows, we will concentrate our theoretical analysis on BRPs against PNS attacks in the typical BB84 protocol.

### III. Optimal eavesdropping scenario and optimal parameters of BRPs

Eve is assumed to have an infinite power obeying the physics laws. In accordance with either single-photon or multi-photon pulses used in QKD, Eve may choose different optimal eavesdropping strategies. For the case of multi-photon pulses, Eve can split at least one photon for probe, which causes no error disturbance to Bob. Unlike the general PNS attack, Eve cannot suppress any single-photon pulses as Bob monitors the BRPs every clock cycle. On the other hand, she cannot increase the yields of multi-photon pulses more than their original yields without splitting attack, because the increase of their yields will betray Eve to Bob. For the case of single-photon pulses, there are two representative scenarios: intercept-resend attack [4] and individual attack [5]. In intercept-resend attack, Eve intercepts the single-photon pulses for measurement. Then she prepares new photon-pulses as her measurement results, and resends them to Bob. In the individual attack, Eve prepares a probe initially in a standard state and lets it interact unitarily with the photon state sent by Alice. She stores the probe until Alice broadcasts her basis, then she measures it. Both scenarios will cause disturbance to Bob even though no additional photon loss is caused in the quantum channel. Eve has a trade-off between the information and the disturbance. According to the analysis in Refs. [4, 5], the individual attack will obtain more information than the intercept-resend attack at the same disturbance. To sum up, Eve can choose the individual attack against the single-photon pulses and the photon-splitting attack against the multi-photon pulses to form her optimal eavesdropping.

We suppose that Bob has high-performance single-photon detectors without any after-pulses. The photon pulses obey the Poissonian distribution $P_n(\mu)=e^{-\mu}\mu^n/n!$. The mean photon numbers of signals and BRPs are $\mu_S$ and $\mu_B$, respectively. The probability that i-photon-number pulses detected by Bob is given by

$$\eta_i = 1-(1-\eta_t\eta_d)^i, \qquad (1)$$

where $\eta_t$ is the channel transmission efficiency ($\eta_t =10^{-0.21L/10}$ for a fiber distance L), and $\eta_d$ is the detecting efficiency of Bob's detectors (including the losses of Bob's decoding components). The overall probability of empty detection of the BRPs is given by

$$G_B(0) = \sum_{i=0}^{\infty} P_i(\mu_B)(1-\eta_t\eta_d)^i = e^{-\eta_t\eta_d\mu_B}. \qquad (2)$$

Note that Eve cannot change Bob's detecting probabilities of signal pulses and BRPs. She can at best suppress $G_B(0)$ of the single-photon number pulses to maximize her information as what she does in the general PNS attack. For a fixed channel transmission efficiency $\eta_t$, one can always find a suitable $\mu_B$ to keep $G_B(0)$ at a negligible value. For the sake of simplicity, we neglect the contribution of $G_B(0)$ to get the best parameters such as the optimal $\mu_\sigma$ and the longest possible transmission distance for the BRPs-QKD.

As proved by Wyner *et. al.* [18], the communication between Alice and Bob can be absolutely secure when Bob's information is more than Eve's, which can be realized by choosing an appropriate data coding. The security criterion can be set as $I_{AB}>I_{AE}$, where $I_{AB}$ and $I_{AE}$ are the mutual information of Alice with Bob and Eve, respectively. The mutual information between

Alice and Bob is given by

$$I_{AB} = 1 + D\log_2(D) + (1-D)\log_2(1-D), \tag{3}$$

where D is Bob's bit error rate and for the sake of simplicity, we only consider all the errors are caused by Eve's disturbance. The mutual information between Alice and Eve includes gains of multi-photon pulses and single-photon pulses. The gain of multi-photon pulses causes no disturbance and therefore Eve can obtain 100% correct information from Alice, which produces a contribution to the mutual information given by

$$I_{AE}(m) = \frac{Y_{\exp} - Y_1}{Y_{\exp}}, \tag{4}$$

where $Y_{exp}$ is Bob's expected click of all the photon pulses, which is given by

$$Y_{\exp} = \sum_0^\infty P_i(\mu_s)\eta_i = 1 - e^{-\eta_t\eta_d\mu_s}. \tag{5}$$

And $Y_1$ is Bob's expected click of single-photon pulses, which is given by

$$Y_1 = P_1(\mu_s)\eta_1 = e^{-\mu_s}\eta_t\eta_d\mu_s. \tag{6}$$

As $\eta_t\eta_d$ is quite small, we can approximately get $Y_{\exp} \approx \eta_t\eta_d\mu_s$ and $Y_1 \approx (\mu_s - P_1(\mu_s))/\mu_s$.

The other gain of the mutual information comes from the single-photon pulses. According to Ref. [5], this part can be estimated by $I_{AE} \leq \frac{1}{2}\phi(2\sqrt{D(1-D)})$, where $\phi(z) = (1+z)\ln(1+z) + (1-z)\ln(1-z)$. As we consider the binary communication in our case, we change $ln$ to $log_2$ in the above expression. The mutual information contributed by the single-photon pulse can then be approximately written as

$$I_{AE}(1) \leq \frac{P_1(\mu_s)}{\mu_s} \cdot (1 + D'\log_2(D') + (1-D')\log_2(1-D')), \tag{7}$$

where $D' = 1/2 - \sqrt{D_{Eve}(1-D_{Eve})}$, and $D_{Eve}$ is the bit error rate of the part of single-photon pulses. Then the total mutual information between Alice and Eve is given by

$$I_{AE} = I_{AE}(1) + I_{AE}(m). \tag{8}$$

Since no disturbance is induced in the quantum channel as Eve obtains the multi-photon pulses, the whole bit error rate comes from the part of single-photon pulses, which is given by

$$D_{Eve} = \frac{\mu_s \cdot D}{P_1(\mu_s)}, \tag{9}$$

We numerically calculate the Eq. 8 with $\mu_s$ varying from 0.1 to 1.0 and obtain the mutual information $I_{AE}$ between Alice and Eve versus Bob's error bit rate ($D$). These results are compared with the mutual information $I_{AB}$ between Alice and Bob calculated by using Eq. 3, as well as $I_{AE}$ for the ideal single-photon source. As shown in Fig.2, the secure bound of $D$ comes close to the case of the ideal single-photon source as $\mu_s$ varies from 1.0 to 0.1. This is expected for the ideal case when we neglect $G_B(0)$, where we only consider the bit error from disturbance induced by Eve. However, this dose not necessarily imply that the small mean photon number of the signals are preferable in a practical QKD application, where Bob's bit error rate is actually determined by some important factors besides the disturbance of Eve. One of the unavoidable bit errors at Bob's site comes from the dark-count and back-ground noises detected by Bob ($Y_{Bob}(0)$). Another factor that contributes to Bob's bit errors in practice is the probability that the photon enters the wrong detector ($e_{detector}$), which is most likely caused by the optical errors, independent of the transmission distance. By taking those factors into account, we can express Bob's bit error rate in the practical QKD application by

$$D_{Bob} = \sum_{i=0}^{\infty} d_i / Y_{\exp} = \sum_{i=0}^{\infty} [d_0 Y_{Bob}(0) + d_{detector} Y_{\exp}] / Y_{\exp} = \frac{e_0 Y_{Bob}(0) + e_{detector}(1 - e^{-\eta_t \eta_d \mu_s})}{1 - e^{-\eta_t \eta_d \mu_s}}, \tag{10}$$

where $e_0$ is the error probability caused by $Y_{Bob}(0)$. As $Y_{Bob}(0)$ randomly takes place, we have $e_0 = 1/2$. $D_{Bob}$ is a function of $\eta_t$ ($L$) and $\mu_s$. From Eq. 6, we find $D_{Bob}$ varies with the transmission distance.

In order to estimate the security of the practical system of a BRPs-QKD, we consider in what follows the dependence of the key generation rate with the transmission distance rather than the mutual information dependence on the bit error rate. The key generation rate is defined by the probability to generate the valid key per pulse, which is determined by not only the mutual efficiency (referring to the bit error rate) but also the transmission efficiency of the channel (referring to the channel loss). The key generation rates of Bob and Eve are given by

$$R_{Bob} = \frac{1}{2} Y_{\exp} I_{AB}, \quad R_{Eve} = \frac{1}{2} Y_{\exp} I_{AE}. \tag{11}$$

The security criterion can be equivalently written as $R_{Bob} > R_{Eve}$. In order to show the capacity of the QKD system, we define the difference between Bob's and Eve's key generation rate as

$$R_S = R_{Bob} - R_{Eve}. \tag{12}$$

Bob can receive secure keys from Alice when $R_S > 0$. Otherwise, their communication is too risk to avoid eavesdropping when $R_S <= 0$. We set the low bound of Bob's key generation rate as $R_{Bob} = R_{Eve}$. Taking as an example, we use the experimental data reported in Ref. [7] to estimate the scenario. The relevant experimental data are as follows. The detection efficiency of Bob's setup is $\eta_d = 0.045$, the bit error rate $e_{detector} \sim 3.3\%$, the loss of fiber 0.21dB/km at 1550 nm, and $Y_{Bob}(0) = 1.7 \times 10^{-6}$. Taking these data into the Eqs. (10-12), we numerically calculate $R_S$ with various $\mu_s$ versus transmission distance. As shown in Fig.3, the optimal value of $\mu_s$ is converged to 0.5 to ensure the longest secure distance up to 146 km and the largest key generation rate at the same transmission distance. This can be understood qualitatively as follows. For the weak pulse with a mean photon number $\mu_s$ used in the practical BRPs-QKD system, the multi-photon probability increases with $\mu_s$, and this part of the information will be leaked to Eve without any disturbance. On the other hand, $D_{Bob}$ is inversely proportional to $\mu_s$, which implies that Eve obtains information of single-photon pulses slowly when $\mu_s$ increases. As a result, $\mu_s=0.5$ is a balance point between these the gains from the multi-photon and single-photon pulses.

After the optimal $\mu_s$ is obtained for the secure BRPs-QKD, we can then estimate $\mu_B$ for its practical implementation. We assume that Eve can only suppress 1/1000 of single-photon pulses. This corresponds to

$$G_B(0) \leq \frac{1}{1000} P_1(\mu_s). \tag{13}$$

And this magnitude of $G_B(0)$ is of no use for Eve. Taking $\mu_s=0.5$ into Eqs. (2) and (13), we find that the low bound of mean photon number $\mu_B$ is exponentially changed with the distance. It should be $2 \times 10^5$ at least for the longest secure distance of 146 km at $\mu_s=0.5$.

### IV. Some improvements for the practical application

The analysis in Sec. III presents the capability of BRPs against PNS attack. If we want to operate the longest distance QKD with the weak pulse source which is secure against the PNS and individual attacks, the best parameters we should choose are $\mu_s=0.5$ and $\mu_B \geq 2 \times 10^5$. As a result, we can implement a 146-km BRPs-QKD. In practice, there still exist two main difficulties to apply BRPs at such a long distance. One is to make perfect interference between signal pulses and bright reference pulses, as $\mu_s$ is much smaller than $\mu_B$. The other is to make sure that Bob will detect the BRP every clock cycle without affecting the measurement of signal pulses.

To create a large ratio as $\mu_B : \mu_s = 2 \times 10^5 : 0.5$, the splitting ratio of BS1 and BS2 should be $2 \times 10^5 : 0.5$. With up-present techniques, the single beam splitter cannot realize such a large ratio either in space or in fiber. However, a combination between a splitter and an attenuator works for this purpose. For example, we may choose experimental parameters as follows for the schematic set-up of the weak-pulse QKD with BRPs in Fig.1. The intensity of the input laser is $8\times10^5$ photons per pulse, the attenuator is -56dB in the short arm of Alice's and Bob's MZ interferometers, and the splitting ratio of BS1 to BS4 is 1:1. The mean photon numbers of the signal pulses and BRPs are $L_l=2\times10^5$ and $L_S=0.5$, respectively. After transmission of a 146-km fiber ($\eta_t=8.6\times10^{-4}$), the corresponding mean photon numbers at Bob's site become $L_{ll}=2\times10^5\eta_t=172$, $L_{l+S}=L_{S+l}=0.5\eta_t=4.3\times10^{-4}$, and $L_{SS}=1.3\times10^{-5}\eta_t=1.1\times10^{-8}$, for BRPs, signal pulses(interference between weak pulses and BRPs), and dim pulses, respectively. This makes the experimental implementation feasible.

It is more difficult to monitor the BRPs without disturbance to the signal measurement. Differing from Ref. [17], the BRPs reach the detectors with a time difference Δt after the signal pulses in our case. Such a time difference can avoid any disturbance to the signal in the same clock cycles. However, the after-pulses caused by the BRPs may tremendously increase the false clicks to the next signals if the avalanche-photodiode based single-photon detectors are used. It may cause large errors at a high rate. For instance, it induces a bit error of 4E-3 per pulse at the repetition rate of 1 MHz when the probability of after-pulses is 0.8% after 1 μs [19]. To avoid after-pulses of the BRPs, it is almost impractical to use the same detectors for the signal pulses. The BRPs should be switched to other detectors. Fortunately, 10-GHz ultra-high-speed optical-switches are commercially available with a crosstalk less than -20dB. Such a low crosstalk induces negligible false clicks for the signal detection. To check the experimental feasibility, we measured the influence of crosstalk on signal detection by inputting sufficient BRP photons and detected photons at Bob's site by using InGaAs avalanche-photodiode based single-photon detectors with the detecting gate at 1 MHz. With about 100 photons per pulse reaching the signle-photon detector at Bob's site, we detected no false clicks for the signal. This simulates the typical above-mentioned experimental situation with a mean photon number of $L_{ll}= 172$ reaching Bob's components after 146-km fibers, under which there should be less than one photon per pulse reach the signal detectors with an optical switch of a crosstalk -20dB. This confirms that the BRPs bring about negligible disturbance to the signal detection in the practical BRP-QKD. The experimental implementation can be further simplified by doubling the intensity of the BRPs so that only one detector is needed to monitor the BRPs after either optical switch of both detection ports as shown in Fig. 1.

### V. Conclusion

In conclusion, we analyze the scenario to beat the photon-number-splitting attacks with the bright reference pulses. We present the optimal eavesdropping as far as we know that Eve use PNS attack against multi-photon pulses and individual attack against single-photon pulses. Afterward we obtain the optimal operation parameters of BRPs with $\mu_s=0.5$ and $\mu_B>2\times10^5$. For the eavesdropping strategy, the BRPs-QKD scenario ensures the secure transmission distance up to 146 km in fiber at 1550 nm. Finally, we present a practical solution to create a large splitting ratio between bright reference and signal pulses for BRPs-QKD, and an experimentally feasible scheme to switch the bright reference pulses away from signal pulses to avoid the after-pulse disturbance. As a result, the BRPs-QKD scenario is highly secure in practical application.

This work was partly supported by Program for Changjiang Scholars and Innovative Research Team in University (PCSIRT), National Natural Science Fund (Grants 10525416 and 10374028), National Key Project for Basic Research (Grant 2001CB309301), key project sponsored by

National Education Ministry of China (Grant 104193), and ECNU PhD Scholarship.

*Information Theory* **24,** 339 (1978)

[19] A. Yoshizawa, R. Kaji, and H. Tsuchida, *Opt. Express* **11**, 1303 (2003)

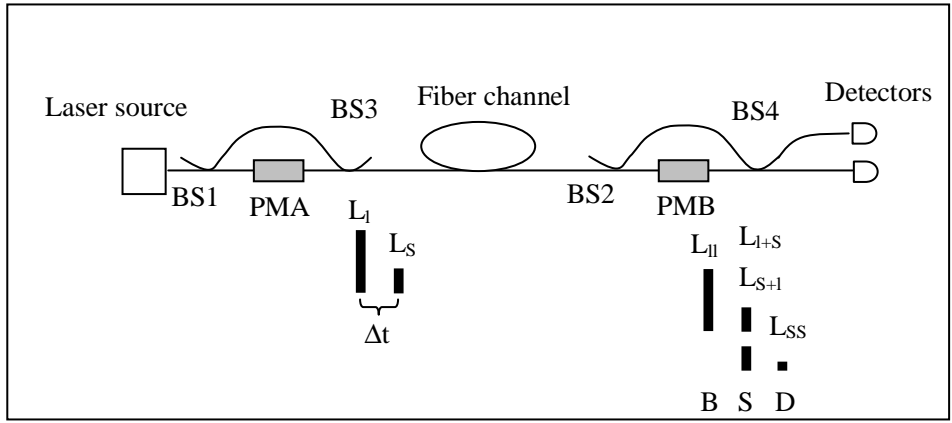

Figure 1, Schematic setup of the weak-pulse QKD with the bright reference pulses, where BS1 and BS2 are the asymmetric beam splitters, BS3 and BS4 are 1:1 beam splitters, PMA and PMB are phase modulators at Alice's and Bob's sites, respectively, and Detectors are used for single-photon detection.

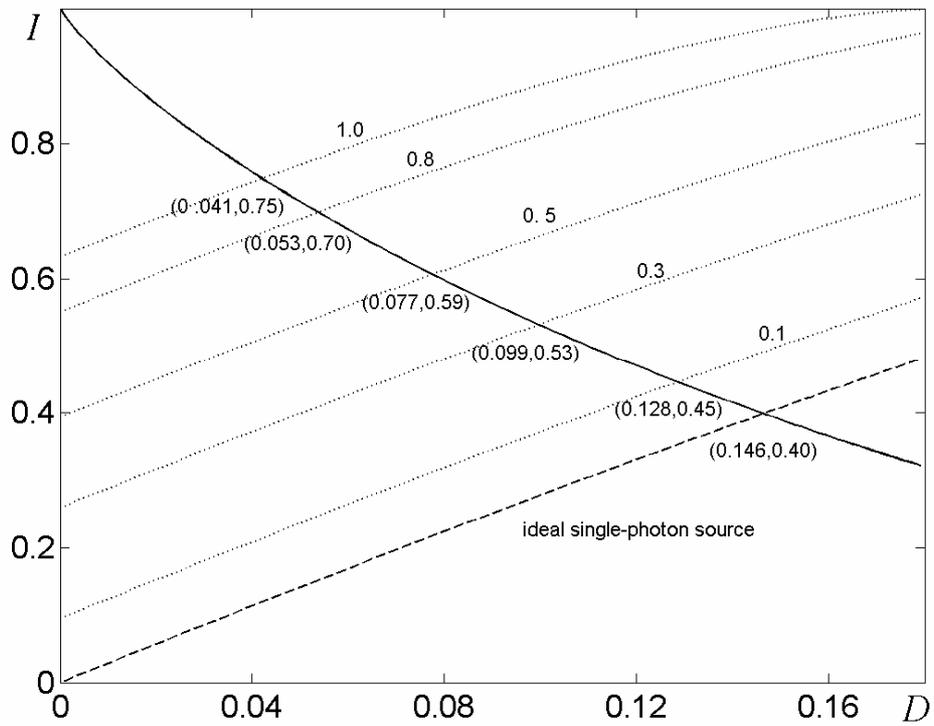

Figure 2, Mutual information (I) vs disturbance (D) for BRPs-QKD with different mean photon number of signal pulses ($\mu_s$), where the solid line is the mutual information between Alice and Bob ($I_{AB}$), the dashed line is $I_{AE}$ with ideal single-photon source, and the dotted lines are $I_{AE}$ of different $\mu_s$.

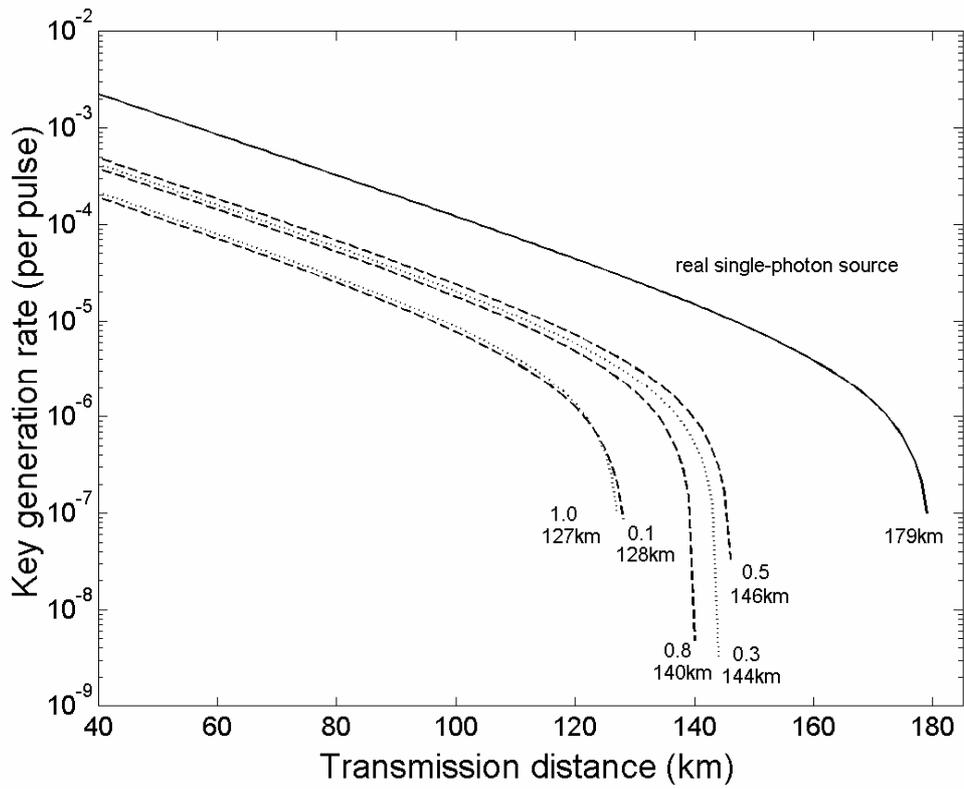

Figure 3, Key generation rate difference ($R_S$) vs transmission distance for BRPs-QKD with different $\mu_s$, where the solid line is $R_S$ with ideal single-photon source, the dashed and dotted lines are $R_S$ with different $\mu_s$.